\documentclass[preprint,12pt]{elsarticle}

\usepackage{amssymb}

\usepackage{bm, color}
\usepackage{amsmath,amsfonts,amssymb,amsthm}
\usepackage{placeins}
\usepackage[title]{appendix}
\usepackage{systeme}
\usepackage{tikz}
\usetikzlibrary{calc}
\usepackage{subfigure}

\usepackage{subfigure}
\usepackage{graphicx}
\usepackage{graphicx}
\usepackage{amsmath,amsfonts,amssymb,bm}
\usepackage{color}
\usepackage{ulem}
\usepackage{float}
\usepackage{caption}
\usepackage{multirow}
\usepackage[overload]{empheq}
\usepackage{cases}
\usepackage{makeidx}         
\usepackage{graphicx}        
\usepackage{multicol}        
\usepackage[bottom]{footmisc}

\newtheorem{definition}{Definition}
\newtheorem{proposition}{Proposition}
\newtheorem{lemma}{Lemma}

\journal{NLS}

\begin{document}

\begin{frontmatter}



\title{Wigner equations for phonons transport and quantum heat flux}


\author{V.D. Camiola$^a$, V. Romano$^a$, G. Vitanza$^a$\\}

\author{$^a$Dipartimento di Matematica e Informatica - Universit\`a degli Studi di Catania}


\begin{abstract}
Starting from the quantum Liouville equation for the density operator and applying the Weyl quantization, Wigner equations for the longitudinal and transversal optical and acoustic phonons are deduced. The equations  are valid for any solid, including 2D crystals like graphene.  With the use of Moyal's calculus and its properties the pseudo-differential operators are expanded up to the second order in $\hbar$. The phonon-phonon collision operators are modelled  in a BGK form and describe the relaxation of the Wigner functions to a local equilibrium function, depending on a local equilibrium temperature which is definite according to \cite{MaRo1}. 

An energy transport model is obtained by using the moment method with closures based on a quantum version of the Maximum Entropy Principle. An explicit form of the thermal conductivity with quantum correction is obtained under a suitable scaling. 

\end{abstract}


%




\end{frontmatter}


\section{Introduction}
The use of the Wigner function is one of the most promising ways to study quantum transport. Its main advantage is that a description similar to the classical or semiclassical transport is obtained in a suitable phase-space.
The mean values are expectation values with respect to the Wigner function as if the latter were a probability density and the semiclassical limit of the Wigner transport equation recovers, at least formally, the Boltzmann transport one. There is a huge body of literature regarding the Wigner equation and the way to numerically solve it (see for example \cite{MoSc, Mu,QuDo} and references therein).
However, the most of the works on the subject consider a quadratic dispersion relation for the
energy. Instead, for several materials like semiconductors or semimetal, e.g. graphene, other dispersion
relations must be considered \cite{Ja,Ju,MaRo2}. From the Wigner transport equation quantum hydrodynamical
models have been obtained in \cite{Ro} for charge transport in silicon in the case of parabolic bands,
while in \cite{LuRo2} the same has been devised for electrons moving in graphene.

The enhanced miniaturization of electron and mechanical devices makes the thermal effects increasingly relevant  \cite{SeCiJou,Marzari} requiring the use of physically accurate models. At kinetic level a good description is that based on the semiclassical Peierls-Boltzmann equation for each phonon branch. However, for typical lengths smaller than the phonon mean-free path also quantum effects must be considered  (see for example \cite{Marzari}).  The Wigner equation is a natural approach that better reveal the wave nature of phonons in such circumstances, gives  the Peierls-Boltzmann equation as semiclassical limit and still keeps the structure of a kinetic formulation. In this work, the focus is on the acoustic and optical phonons dynamics with a general dispersion relation. 

In order to get insights into the quantum corrections,  moment equations are deduced from the corresponding Wigner equation. As in the classical case, one is led to  a system of balance equations that are not closed. So, the well-known problem of getting closure relations arises, that is the issue to express the additional fields appearing in the moment equations in terms of a set of fundamental variables, e.g. the phonon energy density and energy flux.  A sound way 
to accomplish this task is resorting to a quantum formulation of the maximum entropy principle \cite{Jay1} (hereafter QMEP), formulated for the first time by Jaynes \cite{Jay2}. Recently,  a more formal theory has been developed in a series of papers \cite{DeRi,DeMeRi} with several applications, for example for charge transport in semiconductors \cite{Ro,Barletti,BaCi,LuRo3}. The interested reader is also referred to \cite{CaMaRo_book}.  

We apply QMEP to the  Wigner equations assuming  the energy density and the energy flux for each species of phonons as basic fields. By expanding up to the second order in $\hbar$, quantum corrections to the semiclassical case \cite{MaRo1} are deduced. In particular, in a long time scaling an asymptotic expression for the heat flux is   obtained. The latter consists of a Fourier-like part with a highly nonlinear second order correction in the temperature gradient.   Explicit formulas for acoustic phonons in the Debye approximation are written. 

 The plan of the paper is as follows.  In section \ref{PhononsEquations}, the semiclassical phonon transport is summarized while in section \ref{Wigner_equations} we write down the Wigner equations for acoustic and optical phonons. Section \ref{MomentsEquations} is dedicated  to deducing the moment equations whose closure relations are achieved by QMEP in section \ref{sectionQMEP}.  In the last section a definition of local temperature is introduced and an asymptotic expression of the quantum correction to the heat flux is  drawn. 

\section{Semiclassical phonon transport}\label{PhononsEquations}
In a crystal lattice the transport of energy is quantized in terms of quasi-particles named phonons which are present with several branches and propagation modes. The latters vary from a material to another but  in any case they are grouped in acoustic and optical phonon branches which, in turn, can oscillate in the longitudinal or transversal direction. The complete dispersion relations can be usually obtained by a numerical approach in the first Brillouin zone (FBZ) ${\cal B}$. However, in the applications some standard approximations are often adopted.

For the acoustic phonons, the Debye approximation for the dispersion relation ${\bf\varepsilon}_{s}({\bf q})$ is usually assumed, ${\bf\varepsilon}_{s}({\bf q})=\hbar\omega_s({\bf q})=\hbar c_s|{\bf q}|$, $s= LA, TA$. $LA$ stands for longitudinal acoustic while $TA$ for transversal acoustic.  $c_s$ is the sound speed of the s-branch and $\hbar$ denotes the reduced Planck  constant.  Consistently,  the first Brillouin zone is extended to $\mathbb{R}^d$. Here $d$ is the dimension of the space; $d=3$ for bulk crystal while $d=2$ for graphene or similar 2D material like dichalcogenides. 

For the longitudinal  optical (LO) and the  transversal optical (TO) phonon, the Einstein dispersion relation,
$\hbar\omega_s \approx$ const, with $s = LO, TO$, the phonon angular frequency, is usually adopted. 

Note that under such an assumption, the group velocity of the optical phonons is negligible. 

In some peculiar materials like graphene, it is customary to introduce also a fictitious branch called 
$K$-phonons constituted by the phonons having wave vectors close to the Dirac points, $K$ or $K'$, in the first Brillouin zone (taking the origin in the center $\Gamma$ of FBZ). 
Also in this case the Einstein approximation is used on account of the limited variability
of the phonon energy near those points. Moreover, in graphene the phonons are classified as in-plane, representing vibration parallel to the material, and out of plane, representing vibrational mode orthogonal to the material. The $LA$, $TA$, $LO$, $TO$ and $K$ phonons are in plane. 
The out of plane phonons belong to the  acoustic branch and are named $ZA$ phonons. For them a quadratic dispersion relation is a good approximation ${\bf\varepsilon}_{ZA}({\bf q})=\hbar\omega_{ZA}({\bf q})=\hbar\alpha |{\bf q}|^2$, where $\alpha= 6.2\times10^7 m^2/s$ (see \cite{CoRo})

In the following, instead of the wave vector $\bf q$ we will use the phonon moment $\hbar {\bf q}$, but we retain the same character for sake of simplifying the notation. So that ${\bf\varepsilon}_s({\bf q})=c_s|{\bf q}|$, $s=LA,TA$ and ${\bf\varepsilon}_{ZA}({\bf q})=\overline{\alpha}|{\bf q}|^2$, where $\overline{\alpha}=\alpha/\hbar$. 

The thermal transport is usually described by macroscopic models, e.g. the Fourier one, those based on the Maximum Entropy methods \cite{CaMaRo_book} or on phenomenological description \cite{SeCiJou}. A more accurate way to tackle the question is to resort to semiclassical transport equations, the so-called Peierls-Boltzmann equations, 
for each phonon branch for the phonon distributions $f_{\mu} (t,\mathbf{x},\mathbf{q})$ 

\begin{eqnarray}
& &\frac{\partial f_{\mu}}{\partial t} + {\bf c}_{\mu} \cdot  \nabla_{\bf x} f_{\mu} = C_{\mu}, \quad \mu = LA,TA, \ldots,
\end{eqnarray}
where ${\mathbf{c}}_{\mu} = \nabla_{\mathbf{q}}$ $\left(\hbar\omega_{\mu}\right)$ is the group velocity of the $\mu$th phonon specie.  

The phonon collision term $C_{\mu}$ splits into two terms
\begin{eqnarray}
& & C_{\mu} =  C_{\mu}^{\mu}  +  \sum_{\nu, \nu \ne \mu} C_{\mu}^{\nu}, \quad \nu= LA, TA, \ldots.
\end{eqnarray} 
$ C_{\mu}^{\mu}$ describes the phonon interaction within the same branch while $\, C_{\mu}^{\nu}$  describes  the phonon-phonon interaction between different species. To deal with the complete expressions of the $ C_{\mu}$'s  is a very complicated task even from a 
 numerical point of view \cite{Sri}. So,  they are usually simplified by the Bhatnagar-Gross-Krook (BGK) approximation 
 $$ C_{\mu}=   - \frac{f_{\mu} - f_{\mu}^{LE}}{\tau_{\mu} (\mathbf{q})}, $$
 which mimics the relaxation of each phonon branch towards a common local  equilibrium condition, characterised by a local equilibrium  temperature $T_{L}$ that is  the same for each phonon population. 

 The local equilibrium phonon distributions are  given by the Bose-Einstein distributions 
\begin{eqnarray}
& & f_{\mu}^{LE} = \left[\displaystyle{e^{\hbar \omega_{\mu}/k_B T_{L}} } - 1\right]^{-1}. \label{phonons_local_distr}
\end{eqnarray}
and the functions $\tau_{\mu}$ are the phonon relaxation times.
Additional BGK terms can be added to include the interaction between pairs of different branches. 

If we know the phonon distributions $f_{\mu}$'s, we can calculate the average phonon energy densities 
\begin{eqnarray}
& & W_{\mu} = \frac{1}{(2 \pi)^2} \int_{\cal B} \hbar \omega_{\mu} \, f_{\mu} \, d {\bf q},
\end{eqnarray} 
and the expectation value of any  function $\psi(\bf q)$
$$
M_{\psi} = \frac{1}{(2 \pi)^2} \int_{\cal B} \psi({\bf q}) \, f_{\mu} \, d {\bf q},
$$
for example the energy flux if one takes $\psi(\bf q) = \hbar \omega_{\mu} {\bf v}_{\mu}$.

The modern devices, e.g. the electron ones like double gate MOSFETs (see \cite{CaMaRo_book}), are undergoing  more and more miniaturization. This  implies
that the characteristic scales are of the same order as the typical lengths where quantum effects become more and more relevant.  Therefore, quantum effects must be included and the semiclassical phonon transport equations must be replaced by a more accurate model. Among the possible approaches, that one based on the Wigner equation has the advantage to be formulated in a phase-space, allowing us to guess the feature of the solutions in analogy with the semiclassical counterpart.    

A huge literature has been devoted to  the application of the Wigner equations to charge transport (see \cite{MoSc,Mu,QuDo}) but a limited use has been made for phonon transport. In the next sections a transport model, based on the Wigner quasi distribution, will be devised for phonon transport in nano-structures. 

\section{Phonon Wigner functions} \label{Wigner_equations}
The main point of our derivation is the kinetic description of a one-particle quantum
statistical state, given in terms of one-particle Wigner functions. 
 Let us now briefly recall the basic definitions and properties.
A mixed (statistical) one-particle quantum state for an ensemble of scalar particles in
$\mathbb{R}^d$ 
is described by a density operator $\hat{\rho}$, i.e. a bounded non-negative operator with unit trace,
acting on $L^2(\mathbb{R}^d ,\mathbb{C})$. 
Given the density operator $\hat{\rho}$ on $L^2(\mathbb{R}^d ,\mathbb{C})$, 
the associated Wigner function, $w = w(x, \bf q)$, ($x, {\bf q})\in \mathbb{R}^{2d}$, is the inverse Weyl quantization of $\hat{\rho}$,
\begin{equation}
w=Op^{-1}_{\hbar}(\hat{\rho}).
\end{equation}
We recall that the Weyl quantization of a phase-space function (a {\it symbol}) $a = a(x, {\bf q})$ is the (Hermitian) 
operator $Op_{\hbar}(a)$ formally defined by \cite{Hall}
\begin{equation}
Op_{\hbar}(a)\psi(x) = \frac{1}{(2\pi\hbar)^d}\int_{\mathbb{R}^{2d}}a\left(\frac{x+y}{2}, {\bf q}\right)\psi(y)e^{i(x-y)\cdot {\bf q}/\hbar} dy \, d {\bf q}
\end{equation}
for any $\psi \in L^2(\mathbb{R}^d ,\mathbb{C})$.
The inverse quantization of $\hat{\rho}$ can be written  as the {\it Wigner transform}
\begin{equation}
w(x, {\bf q}) =\int_{\mathbb{R}^d}\rho(x+\xi/2, x-\xi/2)e^{i {\bf q}\cdot\xi/\hbar}d\xi,
\end{equation}
of the kernel  $\rho(x, y)$ of the density operator.

The dynamics of the time-dependent phonon Wigner functions $g_{\mu} ({\bf x, q}, t),$ $ {\mu}= LA,TA, \ldots$ steams directly from the dynamics of the corresponding density operator $\hat{\rho}_ {\mu}(t)$, i.e. from the
Von Neumann or quantum Liouville equation 
\begin{equation}\label{VonNeumannac}
i\hbar\partial_t \hat{\rho}_ {\mu}(t) =[ \hat{H}_{\mu},\hat{\rho}_ {\mu}(t)]:=\hat{H}_{\mu} \hat{\rho}_ {\mu}(t)-\hat{\rho}_ {\mu}(t) \hat{H}_{\mu},
\end{equation}
where $\hat{H}_{\mu}$ denotes the Hamiltonian operators of the ${\mu}$th phonons and $[\cdot, \cdot]$ the commutator. If $h_{\mu} = Op^{-1}_{\hbar}(\hat{H}_{\mu})$ is the symbol associated with $\hat{H}_{\mu}$, then, from Eq.s (\ref{VonNeumannac}), we obtain the {\it Wigner equation} for each phonon species
\begin{equation}\label{Wignerac}
i\hbar\partial_t g_{\mu} ({\bf x, q}, t) = \{h_{\mu}, g_{\mu} ({\bf x, q}, t)\}_{\#} := h_{\mu} \# g_{\mu} ({\bf x, q}, t)
 - g_{\mu} ({\bf x, q}, t) \# h_{\mu}. 
\end{equation}
With the symbol $\#$ we have denoted the  Moyal
(or {\it twisted}) product which translates the product of operators at the level of symbols according to
\begin{equation}\label{Moyal}
a \# b = Op^{-1}_{\hbar}(Op_{\hbar}(a)Op_{\hbar}(b)),
\end{equation}
for any pair of symbols $a$ and $b$. Here, we do not tackle the analytical issues which guarantee the existence of the 
previous relations but limit ourselves to the remark that if two operators are in the Hilbert-Schmidt class, that is the trace there exists and it is not negative and bounded,  then the product is still Hilbert-Schmidt and the Moyal calculus is well defined. In the sequel, we will suppose that such conditions are valid. 

The Moyal product, under suitable regularity assumptions (see \cite{Folland}), possesses the following formal semiclassical expansion
\begin{equation}\label{espansion}
a \#_{\hbar}b ({\bf x}, {\bf q})=\sum_{\alpha,\beta}\left(\frac{i\hbar}{2}\right)^{|\alpha|+|\beta|}\frac{(-1)^{|\beta|}}{\alpha!\beta!}\partial_{\bf x}^{\alpha}\partial_{\bf p}^{\beta}a({\bf x}, {\bf q})\partial_{\bf x}^{\beta}\partial_{\bf q}^{\alpha}b({\bf x}, {\bf q})
\end{equation}
where $\alpha=(\alpha_1,...,\alpha_d )\in \mathbb{N}^d$ is a multi-index, $|\alpha|=\sum_i\alpha_i$, $\alpha!=\prod_i\alpha_i!,$ $\partial_{\bf x}^{\alpha}=\prod_i\partial_{x_i}^{\alpha_i}$ and similarly for $\beta$.

The expansion (\ref{espansion}) can be rewritten as
\begin{equation}
a \#_{\hbar}b ({\bf x}, {\bf q}) = \sum_{n=0}^{\infty}\hbar^na\#_nb
\end{equation}
where
\begin{equation}\label{nespansion}
a\#_nb({\bf x}, {\bf q}) = \sum_{\alpha,\beta,|\alpha|+|\beta|=n}\left(\frac{i}{2}\right)^n\frac{(-1)^{|\beta|}}{\alpha!\beta!}\partial_{\bf x}^{\alpha}\partial_{\bf q}^{\beta}a({\bf x}, {\bf q})\partial_{\bf x}^{\beta}\partial_{\bf q}^{\alpha}b({\bf x},{\bf q})
\end{equation}

The first terms of (\ref{nespansion}) read
\begin{eqnarray}
a\#_0b &=& ab,\\
a\#_1b &=& \frac{i}{2}(\nabla_{\bf x} a\cdot\nabla_{\bf q} b-\nabla_{\bf q}a\cdot\nabla_{\bf x} b),\\
a\#_2b &=&-\frac{1}{8}(\nabla_{\bf x}^2a:\nabla_{\bf q}^2b-2\nabla_{\bf x}\nabla_{\bf q}a:\nabla_{\bf q}\nabla_{\bf x}b+\nabla_{\bf q}^2a:\nabla_{\bf x}^2b).
\end{eqnarray}

where $\nabla^2$ denotes the Hessian matrix and $:$ the contracted product of tensors. It is easy to see that
\[
a\#_nb({\bf x}, {\bf q}) =(-1)^nb\#_na({\bf x}, {\bf q}),
\]
that is  the operation $\#_n$ is commutative (respectively anticommutative)
when $n$ is even (respectively  odd).

If we neglect, for the moment,  the phonon-phonon interactions, the Hamiltonian symbol for each phonon branch is given by
\begin{equation}
h_{\mu} ({\bf q})={\bf\varepsilon}_{\mu}({\bf q}) \quad \mu= LA, TA, \ldots.
\end{equation}

By using the Moyal calculus, one can expand the second members of the previous Wigner equations. Up to  first order in $\hbar^2$, we have
\begin{equation}\label{Wigner1ac}
\partial_tg_{\mu}(t) + S[h_{\mu}]g_{\mu}(t) = 0, \quad {\mu} = LA,TA,\ldots,
\end{equation}
where\footnote{Summation over repeated indices is understood from $1$ to $d$.}
\begin{equation}
S[h_{\mu}]g_{\mu}({\bf x}, {\bf q},t):= {\bf c}_{\mu} \cdot \nabla_{\bf x}g_{\mu} ({\bf x}, {\bf q},t) -  \frac{\hbar^2}{24}\frac{\partial_{\bf q}^3h_{\mu}({\bf q})}{\partial q_i\partial q_j \partial q_k}\frac{\partial_{\bf x}^3g_{\mu}({\bf x}, {\bf q},t)}{\partial x_i\partial x_j\partial x_k}+O(\hbar^4))\quad \mu=LA,TA, \cdots.
\end{equation}
%
%
The previous equations describe only ballistic transport and include only the harmonic contribution to the Hamiltonian. In order to describe also  intra and inter-branch phonon-phonon interactions, an additional  anharmonic term $\hat{H}_{int}$ encompassing the high order correction to the Hamiltonian operator must be added. So doing, one  gets the so-called Wigner-Boltzmann equation
\begin{equation}\label{Wigner_coll}
\partial_tg_{\mu}({\bf x}, {\bf q},t) + S[h_{\mu}]g_{\mu}({\bf x}, {\bf q},t) = C_{\mu}({\bf x}, {\bf q},t), \quad {\mu} = LA,TA,\ldots,
\end{equation}
In the quantum case the expression of $C_{\mu}$ is rather cumbersome. For electron transport in semiconductors the interested reader can see \cite{Frommlet}. In certain regimes it is justified to retain the same form of the semiclassical collision operator as the semiclassical case \cite{QuDo}. Here, we adopt a quantum BGK approach and model the collision terms as 
\begin{equation}
C_{\mu} =-\frac{(g_{\mu} - g_{\mu}^{LE})}{\tau_{\mu}({\bf q})},\quad \mu=LA,TA,\ldots. \label{Wigner_collision}
\end{equation}
where
$g_{\mu}^{LE}$ are now Wigner functions of local equilibrium which will be defined later.   

The equation (\ref{Wigner_coll}) along with the expression (\ref{Wigner_collision}) for the collision operator represents our starting point for the phonon transport. 
Note that for the optical phonons under the Einstein approximation for the energy bands one has formally the same transport equation as the semiclassical case because the group velocity vanishes. 

An alternative derivation of (\ref{Wigner_coll}) can be obtained by explicitly writing the von Neumann equation (see \cite{LuRo3,CaMaRo_book} for the details). One obtains
\begin{equation}
S[h_{\mu}] g_{\mu} (t)=\frac{i}{\hbar(2\pi)^d}\int_{\mathbb{R}^d_{\mathbf{x}'}\times\mathbb{R}^d_{{\boldsymbol \nu}}}\left[\varepsilon \left({\bf q} + \frac{\hbar}{2}{\boldsymbol \nu},t\right) - \varepsilon\left({\bf q} - \frac{\hbar}{2} {\boldsymbol \nu}, t\right)\right] g_{\mu} (\mathbf{x}', {\bf q}, t)e^{-i(\mathbf{x}'-\mathbf{x})\cdot {\boldsymbol \nu}}d\mathbf{x}'d {\boldsymbol \nu},\label{S(E)}
\end{equation}
whose expansion is of course in agreement with the Moyal calculus.

\section{Phonon Moment equations}\label{MomentsEquations}

Getting analytical solutions to equations (\ref{Wigner_coll})-(\ref{Wigner_collision}) is a daunting task. Therefore,  viable approaches are numerical solutions based on finite differences or finite elements \cite{MoSc} or stochastic solutions, e.g. those obtained with a suitable modification of the Monte Carlo methods for the semiclassical Boltzmann equation  \cite{Mu}. However, it is possible to have simpler, even if approximate, models  resorting to the moment method for the expectation values of interest. In fact, it is well known that, although not positive definite, the Wigner function is real and the expectation values of an operator can be  formally obtained as an average of the corresponding symbol with respect to $g_{\mu}({\bf x}, {\bf q},t)$. So, for any regular enough weight function $\psi({\bf q})$, 
let us introduce the short notation
\begin{equation}
<\psi>({\bf x}, t) :=\frac{1}{(2\pi)^d}\int_{\mathbb{R}^d} \psi({\bf q})  g_{\mu}({\bf x}, {\bf q},t) d{\bf q},
\end{equation}
which represents a partial average with respect to the phonon moment ${\bf q}$.

More in general, if $a = a({\bf x}, {\bf q})$ is a smooth {\it symbol} then it is possible to prove that the expectation of the (hermitian) operator  
 $A = Op_{\hbar}(a)$ satisfies\footnote{Here we are considering a fixed instant of time.}
 \begin{eqnarray*}
 \mathbb{E}[A] = \mbox{tr} (\hat{\rho} A) = \int_{\mathbb{R}^2d} \rho ({\bf x}, {\bf y}) k_A ({\bf x}, {\bf y}) d {\bf x} d {\bf y} = \frac{1}{(2\pi)^d}\int_{\mathbb{R}^2d} a({\bf x}, {\bf q}) g_{\mu}({\bf x}, {\bf q},t) d{\bf x} d{\bf q}\\ = \int_{\mathbb{R}^d} < a  >  ({\bf x}, t) d{\bf x},
 \end{eqnarray*}
 where $k_A ({\bf x}, {\bf y})$ is the kernel of $A$. 
 
 We want to consider a minimum set of moments whose physical meaning is well clear. In particular, we shall consider the phonon energy and energy flux of each branch
 \begin{eqnarray} 
W_{\mu} ({\bf x}, t) =  <h_{\mu}> ({\bf x}, t), \quad  {\bf Q}_{\mu} ({\bf x}, t) =  < h_{\mu} {\bf c}_{\mu}> ({\bf x}, t).
 \end{eqnarray}  
  Note that the latter is directly related to the heat flux. 
  
The evolution equations for $W_{\mu} ({\bf x}, t)$ and ${\bf Q}_{\mu} ({\bf x}, t)$ are obtained by
multiplying the relative Wigner equation  by $h_{\mu}({\bf q})$, and $h_{\mu}({\bf q}) {\bf c}_{\mu}$ and integrating with respect to ${\bf q}$
\begin{eqnarray}
\begin{array}{l}
\partial_t W_{\mu} ({\bf x}, t) + \displaystyle{\frac{1}{(2\pi)^d} \int_{{\mathbb{R}^d}}} h_{\mu}({\bf q})S[h_{\mu}]g_{\mu}  \, d{\bf q} =  \frac{1}{(2\pi)^d} \displaystyle{\int_{{\mathbb{R}^d}}} h_{\mu}({\bf q})C_{\mu} \, d{\bf q} , \\[0.4cm]
\partial_t {\bf Q}_{\mu} ({\bf x}, t) +  \displaystyle{\frac{1}{(2\pi)^d} \int_{{\mathbb{R}^d}}} h_{\mu}({\bf q}) {\bf c}_{\mu} S[h_{\mu}]g_{\mu}  \, d{\bf q} =  \frac{1}{(2\pi)^d} \displaystyle{\int_{{\mathbb{R}^d}}} h_{\mu}({\bf q}) {\bf c}_{\mu} C_{\mu} \, d{\bf q}.
\end{array}
\quad \mu = LA, TA, \ldots. \label{moment_eqs}
\end{eqnarray}

 We implicitly assume that the resulting integrals there exist, at least in the principal value sense. In order to get some global insight from eq.s (\ref{moment_eqs}), we formally assume
 the following expansions for each phonon branch\footnote{The coefficients of the odd powers in $\hbar$ are assumed zero in according to the previous Moyal expansion.}
 \begin{eqnarray}
 g_{\mu} ({\bf x}, {\bf q},t) = g_{\mu}^{(0)} ({\bf x}, {\bf q},t) + \hbar^2 g_{\mu}^{(2)}({\bf x}, {\bf q},t) + o(\hbar^2).
 \end{eqnarray}

 It is possible to prove, at least formally \cite{Ju}, that  the semiclassical Boltzmann equation is recovered from the Wigner equation as $\hbar \mapsto 0^+$.   Therefore,
 $g_{\mu}^{(0)} ({\bf x}, {\bf q},t)$ can be considered as  the solution $f_{\mu}$ of the semiclassical transport equation.  
Accordingly, we write
\begin{eqnarray}
W_{\mu} = W_{\mu}^{(0)}+\hbar^2 W_{\mu}^{(2)} + o(\hbar^2), \quad {\bf Q}_{\mu}={\bf Q}_{\mu}^{(0)} + \hbar^2 {\bf Q}_{\mu}^{(2)} + o(\hbar^2),
\end{eqnarray}
where
\begin{eqnarray*}
W_{\mu}^{(0)} &=&  \frac{1}{(2\pi)^d} \int_{\mathbb{R}^d} h_{\mu} g_{\mu}^{(0)} ({\bf x}, {\bf q},t) d{\bf q}, \quad W_{\mu}^{(2)} = \frac{1}{(2\pi)^d} \int_{\mathbb{R}^d} h_{\mu} g_{\mu}^{(2)} ({\bf x}, {\bf q},t) d{\bf q},\\
{\bf Q}_{\mu}^{(0)} &=&  \frac{1}{(2\pi)^d} \int_{\mathbb{R}^d} h_{\mu} {\bf c}_{\mu} g_{\mu}^{(0)} ({\bf x}, {\bf q},t) d{\bf q}, \quad {\bf Q}_{\mu}^{(2)} = \frac{1}{(2\pi)^d} \int_{\mathbb{R}^d} h_{\mu} {\bf c}_{\mu} g_{\mu}^{(2)} ({\bf x}, {\bf q},t) d{\bf q}.
\end{eqnarray*}
Regarding the moments of the collision terms, only with drastic simplifications analytical expressions can be deduced. In analogy with the BGK approximation, if an average relaxation time independent on ${\bf q}$ is considered, one can expand the r.h.s. of  eq.s (\ref{moment_eqs}) up to first order in $\hbar^2$ as a relaxation time terms
\begin{eqnarray*}
\frac{1}{(2\pi)^d} \displaystyle{\int_{{\mathbb{R}^d}}} h_{\mu}({\bf q}) C_{\mu} \, d{\bf q} =  -\frac{W_{\mu}-W_{\mu}^{LE}}{\tau_{\mu}^W} = -\frac{W_{\mu}^{(0)}-W_{\mu}^{(0)LE}}{\tau_{\mu}^W}  - \hbar^2 
\frac{W_{\mu}^{(2)}-W_{\mu}^{(2)LE}}{\tau_{\mu}^W} + o(\hbar^2),\\
\frac{1}{(2\pi)^d} \displaystyle{\int_{{\mathbb{R}^d}}} h_{\mu}({\bf q}) {\bf c}_{\mu} C_{\mu} \, d{\bf q} = -\frac{{\bf Q}_{\mu}}{\tau_{\mu}^{\bf Q}}  
= -\frac{{\bf Q}_{\mu}^{(0)} + \hbar^2 {\bf Q}_{\mu}^{(2)}}{\tau_{\mu}^{\bf Q}} + o(\hbar^2),
\end{eqnarray*}
where 
$$
W_{\mu}^{LE} =  \displaystyle{\frac{1}{(2\pi)^d} \int_{{\mathbb{R}^d}}} h_{\mu}({\bf q}) g_{\mu}^{LE}  \, d{\bf q}.
$$
 Note that in the evaluation of the production term of the equations for the energy-fluxes the 
isotropy of the equilibrium Wigner function has been invoked and therefore ${\bf Q}_{\mu}^{LE}$ vanishes.  The energy and energy-flux relaxation times,  $\tau_{\mu}^W$ and $\tau_{\mu}^{\bf Q}$ respectively, are assumed to depends on the temperature, which will be definite in the next section, of the  relative branch.

Altogether, the resulting model is made of the following fluid-type equations
\begin{equation}
\left\{
\begin{array}{l}
\partial_t W_{\mu} + \nabla_{\bf x}\left[{\bf Q}_{\mu}-\dfrac{\hbar^2}{24}\partial^2 {\bf T}_{\mu}\right] = - \displaystyle{\frac{W_{\mu}^{(0)}-W_{\mu}^{(0)LE}}{\tau_{\mu}^W}  - \hbar^2 
\frac{W_{\mu}^{(2)}-W_{\mu}^{(2)LE}}{\tau_{\mu}^W}} + o(\hbar^2)\\
\\
\partial_t {\bf Q}_{\mu} + \nabla_{\bf x}\left[{\bf J}_{\mu}-\dfrac{\hbar^2}{24}\partial^2 {\bf U}_{\mu}\right]=-\displaystyle{\frac{{\bf Q}_{\mu}^{(0)} + \hbar^2 {\bf Q}_{\mu}^{(2)}}{\tau_{\mu}^{\bf Q}}} + o(\hbar^2),\\ 
\end{array}
\right. \label{evoluzione}
\end{equation}
where ${\bf J}_{\mu}={\bf J}_{\mu}^{(0)}+\hbar^2 {\bf J}_{\mu}^{(2)}$
with
\begin{eqnarray*}
& &{\bf J}_{\mu}^{(0)}=\frac{1}{(2\pi)^d}\int_{\mathbb{R}^d}{\bf c}_{\mu}\otimes{\bf c}_{\mu} h_{\mu} ({\bf q})g_{\mu}^{(0)} ({\bf x}, {\bf q},t) d{\bf q} ,\\
& &{\bf J}_{\mu}^{(2)}=\frac{1}{(2\pi)^d}\int_{\mathbb{R}^d}{\bf c}_{\mu}\otimes{\bf c}_{\mu} h_{\mu} ({\bf q})g_{\mu}^{(2)} ({\bf x}, {\bf q},t) d{\bf q},
\end{eqnarray*}
and the complete symmetric tensors ${\bf T}_{\mu}$ and ${\bf U}_{\mu}$ have components
\begin{eqnarray*}
& &({ {\bf T}_{ijk}})_{\mu}=\frac{1}{(2\pi)^d}\int_{\mathbb{R}^d}h_{\mu}\frac{\partial^3 h_{\mu}({\bf q})}{\partial q_i\partial q_j \partial q_k}g_{\mu}^{(0)} ({\bf x}, {\bf q},t) d{\bf q},\\
& &({ {\bf U}_{ijkr}})_{\mu}=\frac{1}{(2\pi)^d}\int_{\mathbb{R}^d}({\bf c}_{\mu})_r h_{\mu}({\bf q})\frac{\partial^3 h_{\mu}({\bf q})}{\partial q_i\partial q_j \partial q_k}g_{\mu}^{(0)}({\bf x}, {\bf q},t) d{\bf q}.
\end{eqnarray*}

If we split into zero and first order in $\hbar^2$, the evolution equations read 

\begin{eqnarray}
&&\partial_t W_{\mu}^{(0)} + \nabla_{\bf x} {\bf Q}_{\mu}^{(0)} = -\frac{W_{\mu}^{(0)}-W_{\mu}^{(0)LE}}{\tau_{\mu}^W} \label{evolution1}\\
&&\partial_t W_{\mu}^{(2)} + \nabla_{\bf x} {\bf Q}_{\mu}^{(2)}+\frac{1}{(2\pi)^d}\frac{\partial^3}{\partial x_i\partial x_j \partial x_k}\int_{\mathbb{R}^d} \frac{h_{\mu}({\bf q})}{24}  g_{\mu}^{(0)} \frac{\partial^3}{\partial q_i\partial q_j \partial q_k} h_{\mu}({\bf q}) d{\bf q}\nonumber\\
& &  = - \frac{W_{\mu}^{(2)}-W_{\mu}^{(2)LE}}{\tau_{\mu}^W}, \\
&&\partial_t {\bf Q}_{\mu}^{(0)} + \nabla_{\bf x}{\bf J}_{\mu}^{(0)} = -\displaystyle{\frac{{\bf Q}_{\mu}^{(0)}}{\tau_{\mu}^{\bf Q}}}, \\
&&\partial_t {\bf Q}_{\mu}^{(2)} + \nabla_{\bf x}{\bf J}_{\mu}^{(2)}+\frac{1}{(2\pi)^d}\frac{\partial^3}{\partial x_i\partial x_j \partial x_k}\int_{\mathbb{R}^d}{\bf c}_s\frac{h_{\mu}({\bf q})}{24} g_{\mu}^{(0)} \frac{\partial^3}{\partial q_i\partial q_j \partial q_k} h_{\mu}({\bf q})d{\bf q} = -\displaystyle{\frac{{\bf Q}_{\mu}^{(2)}}{\tau_{\mu}^{\bf Q}}}. \quad \label{evolution4}
\end{eqnarray}
The zero order equations are the model already investigated in several papers \cite{MaRo1,MaRo2} where is proved that it is a hyperbolic system of conservation law. So, finite propagation speed of disturbances in energy is guaranteed to overcome the well-known paradox of the classical Fourier law for the heat flux. However, the first order corrections in $\hbar^2$  
introduce dispersive terms and it seems that in a quantum regime the requirement of finite propagation speed of the thermal effects  cannot be fulfilled. On the other hand, this is not surprising since the nonlocal character of the quantum evolution equations can lead to energy propagation without a bounded speed.

\section{QMEP for the closure relations}\label{sectionQMEP}
The evolution equations (\ref{evolution1})-(\ref{evolution4}) do not form a closed system of balance laws. If we assume the energies $W_{\mu}$ and the energy-fluxes ${\bf Q}_{\mu}$ as the main fields, in order to get a set of closed equations we need to express the additional fields ${\bf J}_{\mu}$, ${\bf T}_{\mu}$ and ${\bf U}_{\mu}$ as functions of 
$W_{\mu}$ and  ${\bf Q}_{\mu}$. 
A successful approach in a semiclassical setting is that based on the Maximum Entropy Principle (MEP) (see also \cite{CaMaRo_book} for a complete review) which is based on a pioneering paper of Jaynes \cite{Jay1,Jay2} who also proposed a way to extend the approach to the quantum case.
The MEP in a quantum setting has been the subject of several papers \cite{Ro,DeRi,DeMeRi,Barletti,BaCi} with several applications, e.g. to charge transport in graphene \cite{MaRo1,LuRo3}. Here we will use such an approach for phonon transport.  

The starting point is the entropy for the quantum system under consideration. 
In \cite{LuRo3} the authors have employed the Von-Neumann entropy which, however, does not take into account the statistical aspects. Therefore, we take as entropy a generalization of the classical one for bosons. Let us introduce the operator 
\begin{equation}\label{entropy}
s(\hat{\rho}_{\mu}) = -k_B[\hat{\rho}_{\mu} \ln \hat{\rho}_{\mu}-(1+ \hat{\rho}_{\mu}) \ln(1+ \hat{\rho}_{\mu})],
\end{equation}
which must be intended in the sense of the functional calculus. Here $k_B$ is the Boltzmann constant.
The entropy of the ${\mu}$-th phonon branch reads 
$$S(\hat{\rho}_{\mu})=\mbox{Tr} \{s(\hat{\rho}_{\mu})\}$$
which can be viewed as a quantum Bose-Einstein entropy.

According to MEP, we estimate $\hat{\rho}_{\mu}$ with $\hat{\rho}^{MEP}_{\mu}$ which is obtained by maximizing $S(\hat{\rho}_{\mu})$ under the constraints that some expectation values have to be preserved.  In the semiclassical point case,  one maximizes the entropy   preserving the values of the moments we have taken as basic field variables
\begin{equation}\label{moments}
(W_{\mu}({\bf x}, t), {\bf Q}_{\mu}({\bf x}, t))= \dfrac{1}{(2\pi)^d} \int_{\mathbb{R}^d}\bm{\psi}_{\mu}{({\bf q})}g_{\mu}({\bf x,q},t)d{\bf q}  =
 \dfrac{1}{(2\pi)^d} \int_{\mathbb{R}^d}\bm{\psi}_{\mu}{({\bf q})}g^{MEP}_{\mu}({\bf x}, {\bf q},t) d{\bf q},
\end{equation}
where
\begin{equation}\label{weights}
\bm{\psi}_{\mu}{({\bf q})}=(h_{\mu}({\bf q}),{\bf c}_{\mu} h_{\mu}({\bf q}))
\end{equation}
is the vector of the weight functions
and $g^{MEP}_{\mu}$ is the Wigner function associated with $\hat{\rho}^{MEP}_{\mu}$. In the previous relations the time $t$ and position ${\bf x}$ must be considered as fixed. 

The quantum formulation of MEP is given  in terms of expectation values 
\begin{eqnarray*} 
E_1 (t) = \mbox{tr}\left\{ \hat{\rho}_{\mu} Op_{\hbar} (h_{\mu}({\bf q}))\right\} (t), \quad
 {\bf E}_2 (t) =  \mbox{tr}\left\{ \hat{\rho}_{\mu} Op_{\hbar} ({\bf c}_{\mu}  h_{\mu}({\bf q}))\right\}(t),
\end{eqnarray*}
as follows: for fixed $t$
\begin{eqnarray}
& &\hat{\rho}^{MEP}_{\mu} = \mbox{argument max} \, S(\hat{\rho}_{\mu}) \label{QMEP1}\\
& &\mbox{ under the constraints}\nonumber\\
& &\mbox{tr}\{ \hat{\rho}^{MEP}_{\mu} Op_{\hbar}  (h_{\mu}({\bf q})) \} = E_1 (t), \quad 
\mbox{tr}\{ \hat{\rho}^{MEP}_{\mu} Op_{\hbar} ({\bf c}_{\mu}  h_{\mu}({\bf q})) \} =  {\bf E}_2 (t), \label{QMEP2}
\end{eqnarray}
in the space of the Hilbert-Schmidt operators on $L^2(\mathbb{R}^d ,\mathbb{C})$ which are positive, with trace one and such that the previous expectation values there exist. Note that we are applying the maximization of the entropy for each phonon branch separately. In other words, we are requiring the additivity of the entropy.

If we introduce the vector of the Lagrange multipliers
\begin{equation}\label{multipliers}
\bm{\eta}_{\mu} = (\eta_{0 \mu}({\bf x}, t),\bm{\eta}_{1 \mu}({\bf x}, t)),
\end{equation}
the vector of the moments 
\begin{equation}\label{moments}
{\bf m[\rho_{\mu}](x,t)}:={\bf m}_{\mu}({\bf x}, t) =  \dfrac{1}{(2 \pi)^d}\int_{\mathbb{R}^d}\bm{\psi}_{\mu} {({\bf q})}g_{\mu} ({\bf x,q},t)d{\bf q},
\end{equation}
and the vector of the  moments which must be considered as known
\begin{equation} \label{expectations}
{\bf M}_{\mu}({\bf x}, t):=\left(W_{\mu}({\bf x}, t), {\bf Q}_{\mu}({\bf x}, t)  \right),
\end{equation} 
the constrained optimization problem (\ref{QMEP1})-(\ref{QMEP2}) can be rephrased as a saddle-point problem for the Lagrangian
\begin{eqnarray}\label{objective}
{\cal L}_{\mu}(\hat{\rho}_{\mu},\bm{\eta}_{\mu}) &=& S(\hat{\rho}_{\mu}) - \int_{\mathbb{R}^d} \bm{\eta}_{\mu}  \cdot \left(
 {\bf m}_{\mu}({\bf x}, t) - {\bf M}_{\mu}({\bf x}, t) \right) \, d {\bf x} 
\nonumber\\
 & =& S(\hat{\rho}_{\mu}) - 
\mbox{tr}\left\{ \hat{\rho}_{\mu} Op_{\hbar} (\bm{\eta}_{\mu} \cdot h_{\mu}({\bf q}),{\bf c}_{\mu} h_{\mu}({\bf q}))\right\}  + \int_{\mathbb{R}^d} \bm{\eta}_{\mu}  \cdot  {\bf M}_{\mu}({\bf x}, t)  \, d {\bf x} 
\end{eqnarray}
in the space of the admissible $\hat{\rho}_{\mu}$ and smooth function $\bm{\eta}_{\mu}$.


If the Lagrangian  ${\cal L}_{\mu}(\hat{\rho}_{\mu},\bm{\eta}_{\mu})$ is G\^ateaux-differentiable with respect to $\hat{\rho}_{\mu}$, the first order optimality conditions require
$$
\delta {\cal L}_{\mu}(\hat{\rho}_{\mu}, \bm{\eta}_{\mu}) (\delta \hat{\rho}) = 0 
$$
for each Hilbert-Schmidt operators $\delta \hat{\rho}$  on $L^2(\mathbb{R}^d ,\mathbb{C})$  which is positive, with trace one and such that the previous expectation values there exist.

The existence of the first order G\^ateaux derivative is a consequence of the following Lemma (for the proof see \cite{Nier}; an elementary proof in the case of discrete spectrum is given in  \cite{DeRi}).
\begin{lemma} \label{lemma1}
If $r(x)$ is a continuously differentiable increasing function on $\mathbb{R}^+$ then $\mbox{tr} \{ r(\hat{\rho})\}$ is G\^ateaux-differentiable in the class of the Hermitian Hilbert-Schmidt positive operators  on $L^2(\mathbb{R}^d ,\mathbb{C})$. The G\^ateaux derivative along 
$\delta \rho$ is given by
\begin{equation}
\delta \mbox{tr} \{ r(\hat{\rho})\} (\delta \hat{\rho}) = \mbox{tr} \left\{r'(\hat{\rho}) \delta \hat{\rho} \right\}.
\end{equation}
\end{lemma}
The extremality conditions for the unconstrained minimization problem (\ref{QMEP1})-(\ref{QMEP2}) are similar to that of the semiclassical case, as expressed by the following lemma (see \cite{DeRi}).
\begin{lemma}
The first order optimality condition for the  minimization problem (\ref{QMEP1})-(\ref{QMEP2}) is equivalent to 
\begin{equation}
\hat{\rho}_{{\mu}}=(s')^{-1}(Op_{\hbar}(\bm{\eta}_{\mu} \cdot \bm{\psi}_{\mu}))
\end{equation}
where $(s')^{-1}$ is the inverse function of the first derivative of $s$.
\end{lemma}
{\sf Proof}. By applying Lemma \ref{lemma1}, the G\^ateaux derivative of the Lagrangian is given by
$$
\delta {\cal L}_{\mu}(\hat{\rho}_{\mu}, \bm{\eta}_{\mu}) (\delta \hat{\rho}) = \mbox{tr} \left\{ \left(s'(\hat{\rho}_{\mu}) - Op_{\hbar}(\bm{\eta}_{\mu} \cdot \bm{\psi}_{\mu})\right) \delta \hat{\rho} \right\}
$$ 
$\forall \delta \hat{\rho}$ perturbation in the class of the Hermitian Hilbert-Schmidt positive operators  on $L^2(\mathbb{R}^d ,\mathbb{C})$. This implies
$$
s'(\hat{\rho}_{\mu}) = Op_{\hbar}(\bm{\eta}_{\mu} \cdot \bm{\psi}_{\mu}).
$$
\hfill $\Box$

Since the function $s(x)$ is concave, $s'(x)$ is invertible. Explicitly 
we have
\begin{equation*}
(s')^{-1}(z)=\frac{1}{e^{z/k_B}-1}
\end{equation*}
and  the operator solving the first order optimality condition reads
\begin{equation}
\hat{\rho}^*_{\mu} = (s')^{-1}(Op_{\hbar}(\bm{\eta}_{\mu} \cdot \bm{\psi}_{\mu}))=\frac{1}{e^{Op_{\hbar}(\bm{\eta}_{\mu} \cdot \bm{\psi}_{\mu})}-1}.
\end{equation}
Moreover, such an operator is a point of maximum for the Lagrangian.
\hfill $\Box$
\vskip 0.4cm

Now, to complete the program  we have to determine, among  the smooth functions, the Lagrange multipliers  $\bm{\eta}_{\mu}$ by solving the constraint
\begin{equation} 
\mbox{tr}\left\{ \hat{\rho}_{\mu} Op_{\hbar} (\bm{\eta}_{\mu} \cdot ( h_{\mu}({\bf q}),{\bf c}_{\mu} h_{\mu}({\bf q}))\right\} - \int_{\mathbb{R}^d} \bm{\eta}_{\mu}  \cdot  {\bf M}_{\mu}({\bf x}, t)  \, d {\bf x} = 0. \label{vincolo_MEP}
\end{equation}
If such an equation has a solution $\bm{\eta}^*_{\mu}$,
altogether the MEP density operator reads
\vskip 0.2cm
\begin{equation}
\hat{\rho}^{MEP}_{\mu} = 
         \frac{1}{\exp \left[Op_{\hbar}\left( \eta^*_{0 \mu}({\bf x},t)h_{\mu}({\bf q})+\bm{\eta}^*_{1 \mu}({\bf x},t) \cdot {\bf c}_{\mu} h_{\mu}({\bf q})\right)  \right] -1},\label{solution}
\end{equation}
\vskip .2cm
\noindent where we have rescaled the Lagrange multipliers including the factor $1/k_B$. 

To determine conditions under which the equation (\ref{vincolo_MEP}) admits solutions is a very difficult task. Even in the semiclassical case
there are examples (see \cite{Junk}) of sets of moments that cannot be moments of a MEP distribution.  
We will directly find out the solution at least up to first order in $\hbar^2$.

Once the MEP density function has been determined, the MEP Wigner function is given by 
$$g^{MEP}_{\mu}({\bf x}, {\bf q},t) = Op_{\hbar}^{-1}(\hat{\rho}^{MEP}_{\mu})$$
which can be used to get the necessary closure relations by evaluating the additional fields  with $g_{\mu}$ replaced by $g_{\mu}^{MEP}$.

We remark that the constraints (\ref{vincolo_MEP}) can be more conveniently expressed as 
$$
\dfrac{1}{(2\pi)^d}  \int_{\mathbb{R}^2d} \bm{\eta}_{\mu}  \cdot  \bm{\psi}_{\mu}({\bf x}, t) g_{\mu}^{MEP}({\bf x}, {\bf q}, t) \, d {\bf q} \, d {\bf x} - \int_{\mathbb{R}^d} \bm{\eta}_{\mu}  \cdot  {\bf M}_{\mu}({\bf x}, t)  \, d {\bf x} = 0
$$
and indeed we will require, in analogy with the semiclassical case, the stronger conditions
$$
\dfrac{1}{(2\pi)^d}  \int_{\mathbb{R}^d}  \bm{\psi}_{\mu}({\bf x}, t) g_{\mu}^{MEP}({\bf x}, {\bf q}, t) \, d {\bf q}  = {\bf M}_{\mu}({\bf x}, t),
$$
where the Lagrange multipliers enter through  $g_{\mu}^{MEP}({\bf x}, {\bf q}, t)$.

\subsection{Determination of the Lagrange Multipliers}
For the sake of making lighter the notation, let us consider a single branch and drop the index $\mu$ in the Wigner function in this section. We look formally for a solution in powers of $\hbar$ 
\begin{equation}
g^{MEP} = g^{MEP}_0 +\hbar g^{MEP}_1 +\hbar^2 g^{MEP}_2 + ... 
\end{equation}
firstly without taking into account the dependence of the Lagrange multipliers on $\hbar$.

Of course, on account of the properties of the Weyl quantization, $g^{MEP}_0$ is equal to the semiclassical counterpart \cite{Hall}
$$
g^{MEP}_0 =  \frac{1}{\exp \left[ \eta_{0}({\bf x},t) h({\bf q})+ \bm{\eta}_{1}({\bf x},t) \cdot {\bf c} h({\bf q}) \right] -1}
$$
In order to determine the higher order terms $g^{MEP}_k$, $k \ge 1$, 
given a symbol $a({\bf x}, {\bf q})$ let us introduce the so-called {\it quantum exponential} $Exp(a)$ defined as 
 $$Exp(a) = Op^{-1}_{\hbar} [exp(Op_{\hbar}(a))]$$ 
 which can be expanded as
 \begin{equation}
Exp(a) = Exp_0(a)+\hbar Exp_1(a) +\hbar^2 Exp_2(a) + ...
\end{equation}
{\bf Proposition}
Let $a({\bf x},{\bf p})$ be a smooth symbol. Then the following expansion is valid
\begin{eqnarray}
Exp(a) = \exp (a) -\dfrac{\hbar^2}{8}
\exp(a)\left(\frac{\partial^2 a}{\partial{x_i}\partial{x_j}} \frac{\partial^2 a}{\partial{p_i}\partial{p_j}}-\frac{\partial^2 a}{\partial{x_i}\partial{p_j}}\frac{\partial^2 a}{\partial{p_i}\partial{x_j}}
+\frac 13 \frac{\partial^2 a}{\partial{x_i}\partial{x_j}} \frac{\partial a}{\partial{p_i}}\frac{\partial a}{\partial{p_j}} \right.\nonumber\\
\left. - \frac 23 \frac{\partial^2 a}{\partial{x_i}\partial{p_j}}\frac{\partial a}{\partial{p_i}}\frac{\partial a}{\partial{x_j}}  
 + \frac13\frac{\partial^2 a}{\partial{p_i}\partial{p_j}}\frac{\partial a}{\partial{x_i}}\frac{\partial a}{\partial{x_j}}\right)+O(\hbar^4), \label{exp}
\end{eqnarray}
where Einstein's convention has been used.
\hfill $\Box$\\
The proof can be found for example in \cite{DeRi}.

By using what is proved in \cite{Barletti}, we have
\begin{subequations}
\begin{equation}
g^{MEP}_{2n +1} = 0,\quad n\geq 0,
\end{equation}
\begin{equation}
g^{MEP}_{2n} =-\sum_{m=0}^{n-1}\sum_{k+l+m=n}\frac{Exp_{2k}(\xi) \#_{2l} g^{MEP}_{2m}}{e^\xi-1}, \quad n\geq1 
\end{equation}
\end{subequations}
where $\#_{2l}$ are the even terms of the Moyal product expansion and 
$$\xi =\eta_{0 \mu}({\bf x},t) h({\bf q})+ \bm{\eta}_{1}({\bf x},t) \cdot {\bf c} h({\bf q}).$$

In particular 
$$
g^{MEP}_{1} = 0
$$
and 

\begin{eqnarray*}
g^{MEP}_{2} &=& -\frac{1}{8}\frac{e^\xi}{(e^\xi-1)^3}\left[(e^\xi+1)\left(\frac{\partial^2 \xi}{\partial x_i\partial x_j}\frac{\partial^2\xi}{\partial q_i\partial q_j}-\frac{\partial^2 \xi}{\partial x_i\partial q_j}\frac{\partial^2 \xi}{\partial q_i\partial x_j}\right)\right.\\
& &\left.-\frac{(e^{2\xi}+4e^\xi+1)}{3(e^\xi-1)}\left(\frac{\partial^2 \xi}{\partial x_i\partial x_j}\frac{\partial \xi}{\partial q_i}\frac{\partial \xi}{\partial q_j}-2\frac{\partial^2 \xi}{\partial x_i\partial q_j}\frac{\partial \xi}{\partial q_i}\frac{\partial \xi}{\partial x_j}+\frac{\partial^2 \xi}{\partial q_i\partial q_j}\frac{\partial \xi}{\partial x_i}\frac{\partial \xi}{\partial x_j}\right)\right]
\end{eqnarray*}

Therefore, up to first order in $\hbar^2$ we have
\[
g_{\mu}^{MEP} =  g_{0}^{MEP} +  \hbar^2 g_{2}^{MEP}.
\]
and the constraints  for each phonon branch read
\begin{eqnarray}
\label{energy}
 W &=& \frac{1}{(2\pi)^d}\int_{\mathbb{R}^d}\frac{h({\bf q})}{e^\xi-1} d{\bf q}+\hbar^2\frac{1}{(2\pi)^d}\int_{\mathbb{R}^d} h({\bf q}) g_{2}^{MEP} d{\bf q},\\[0.3cm]
{\bf Q} &=& \frac{1}{(2\pi)^d}\int_{\mathbb{R}^d}\frac{ {\bf c} h({\bf q})}{e^{\xi} -1}d{\bf q}+\hbar^2\frac{1}{(2\pi)^d}\int_{\mathbb{R}^d} {\bf c}  h({\bf q}) g_{2}^{MEP} d{\bf q}. \label{flux}
\end{eqnarray}
The previous equations form a nonlinear system of PDEs for the Lagrange multipliers whose analytical solution seems very difficult to get. Indeed, the situation is even more cumbersome because in a numerical scheme the inversion of the constraints should be performed at each time step. 

A viable strategy is to use the Lagrange multipliers as field variables by rewriting the evolution equations (\ref{evoluzione}) in the form
\begin{eqnarray}
&&\nabla_{\bm{\eta}} W  \frac{\partial }{\partial t} \bm{\eta}^T +  \sum_{i = 1}^d  \left[\nabla_{\bm{\eta}} Q_i  \frac{\partial }{\partial x_i} \bm{\eta}^T 
- \dfrac{\hbar^2}{24} \nabla_{\bm{\eta}} \left(\nabla_{\bf x}\partial^2_{\bf x}{\bf T} \right) \frac{\partial }{\partial x_i} \bm{\eta}^T \right]
= -\frac{W -W^{LE}}{\tau^W},\\
&&\nabla_{\bm{\eta}} Q_i  \frac{\partial }{\partial t} \bm{\eta}^T +  \sum_{j = 1}^d \left[
 \nabla_{\bm{\eta}}{\bf J} \frac{\partial }{\partial x_j} \bm{\eta}^T -\dfrac{\hbar^2}{24} \nabla_{\bm{\eta}}\left( \partial^2_{\bf x}{\bf U}\right)\frac{\partial }{\partial x_j} \bm{\eta}^T \right]=-\displaystyle{\frac{Q_i}{\tau^{\bf Q}}},
\end{eqnarray}
getting a highly nonlinear system of PDEs. Note that both $\nabla_{\bm{\eta}} W$ and $\nabla_{\bm{\eta}} Q_i$ contain space derivatives of 
$\bm{\eta}$.

A further simplification can be obtained by expanding the Lagrange multipliers as
$$
\bm{\eta} = \bm{\eta}^{(0)} + \hbar^2 \bm{\eta}^{(2)} + o(\hbar^2).
$$
Therefore, the basic fields are also expanded with respect to $\hbar^2$  
$$
W = W^{(0)} + \hbar^2 W^{(2)} + o(\hbar^2), \quad {\bf Q} = {\bf Q}^{(0)} + \hbar^2 {\bf Q}^{(2)} + o(\hbar^2)
$$
where
\begin{eqnarray*}
\label{energyh}
 W^{(0)} &=& \frac{1}{(2\pi)^d}\int_{\mathbb{R}^d}\frac{h({\bf q})}{e^{\xi^{(0)}}-1} d{\bf q},\\
 W^{(2)} &=&   -  \frac{1}{(2\pi)^d} \bm{\eta}^{(2)} \cdot \int_{\mathbb{R}^d} e^{\xi^{(0)}} \frac{h({\bf q}) \bm{\psi}}{\left(e^{\xi^{(0)}}-1\right)^2} d{\bf q} 
 + \frac{1}{(2\pi)^d}\int_{\mathbb{R}^d} h({\bf q}) g_{2}^{MEP} (\bm{\eta}^{(0)}) d{\bf q},\\
 {Q}^{(0)}_i  &=& \frac{1}{(2\pi)^d}\int_{\mathbb{R}^d}\frac{ c_i h({\bf q})}{e^{\xi^{(0)}} -1}d{\bf q},\\
 {Q}^{(2)}_i&=& - \ \frac{1}{(2\pi)^d} \bm{\eta}^{(0)} \cdot \int_{\mathbb{R}^d}\frac{ c_i  \bm{\psi} e^{\xi^{(0)}} h({\bf q})}{(e^{\xi^{(0)}} -1)^2}d{\bf q}+\frac{1}{(2\pi)^d}\int_{\mathbb{R}^d}  c_i h({\bf q})  g_{2}^{MEP}  (\bm{\eta}^{(0)})  d{\bf q}, \label{flux}
 \end{eqnarray*}
 with $\xi^{(0)} = \bm{\eta}^{(0)} \cdot \bm{\psi}$.
 
The balance equations become
\begin{eqnarray}
&&\nabla_{\bm{\eta}^{(0)}}  W^{(0)} \frac{\partial }{\partial t} (\bm{\eta}^{(0)})^T+ \sum_{i = 1}^d  \left[\nabla_{\bm{\eta}^{(0)}}  {\bf Q}^{(0)} 
\frac{\partial }{\partial x_i} (\bm{\eta}^{(0)})^T \right] = -\frac{W^{(0)}-W^{(0)LE}}{\tau^W} \label{evolution1h}\\
&&\nabla_{\bm{\eta}^{(0)}} Q_i^{(0)}  \frac{\partial }{\partial t} (\bm{\eta}^{(0)})^T + \sum_{i = 1}^d \left[
 \nabla_{\bm{\eta}^{(0)}}{\bf J}^{(0)} \frac{\partial }{\partial x_j} (\bm{\eta}^{(0)})^T \right]  = -\displaystyle{\frac{{Q}^{(0)}_i}{\tau^{\bf Q}}}, \label{evolution2h}\\
&&\partial_t W^{(2)} + \nabla_{\bf x} {\bf Q}^{(2)}+\frac{1}{(2\pi)^d}\frac{\partial^3}{\partial x_i\partial x_j \partial x_k}\int_{\mathbb{R}^d} \frac{h ({\bf q})}{24}  g_{0}^{MEP} (\bm{\eta}^{(0)}) \frac{\partial^3}{\partial q_i\partial q_j \partial q_k} h({\bf q}) d{\bf q} \nonumber\\
 & & = - \frac{W^{(2)}-W^{(2)LE}}{\tau^W}, \label{evolution3h}\\
&&\partial_t {\bf Q}^{(2)} + \nabla_{\bf x}{\bf J}^{(2)}+\frac{1}{(2\pi)^d}\frac{\partial^3}{\partial x_i\partial x_j \partial x_k}\int_{\mathbb{R}^d}{\bf c}\frac{h ({\bf q})}{24} g_{0}^{MEP} (\bm{\eta}^{(0)})  \frac{\partial^3}{\partial q_i\partial q_j \partial q_k} h({\bf q})d{\bf q} = -\displaystyle{\frac{{\bf Q}^{(2)}}{\tau^{\bf Q}}}. \qquad\label{evolution4h}
\end{eqnarray}

We observe that the equations (\ref{evolution1h})-(\ref{evolution2h}) decouple. Once they are solved, one can get the second order term of the Lagrange multipliers from (\ref{evolution3h})-(\ref{evolution4h}) which form a linear system for $\bm{\eta}^{(2)}$. This is rather beneficial from a computational point of view
\begin{proposition}
At zero order in $\hbar^2$ the map 
$
{\bm \eta} \mapsto {\bf M}({\bm \eta})
$
is (locally) invertible.
\end{proposition}
\begin{proposition}
The equations (\ref{evolution1h})-(\ref{evolution2h}) form a symmetric hyperbolic system of balance laws.
\end{proposition}
The proofs can be found in \cite{CaMaRo_book}.

\section{Local equilibrium temperature and heat conductivity}\label{LocalTemperature}
The concept of temperature out of equilibrium is a subtle topic and still a matter of debate. In the case of charge transport in semiconductors often the phonons are considered as a thermal bath and under some reasonable assumptions one can hypothesize that the electrons are in thermal equilibrium with the bath. In general if the dynamics of the phonons must be included, a thermal bath for these does not exist, unless a thermostated system is considered. Therefore, we need to introduce a local equilibrium temperature  for the overall phonon system. 

In statistical mechanics, one of the most reasonable and adopted ways to generalize the concept of temperature in a non-equilibrium state is that of relating it to the Lagrange multipliers associated with the energy constraint. For the phonon transport in graphene, an approach based on the Lagrange multipliers was followed in \cite{MaRo1} (which the interested reader is referred to for the details).  Let us recall here the main features.
At equilibrium, the phonon temperatures and  the corresponding Lagrange multipliers are related by 
\[
k_B \, T_\mu ({\bf x}) =\frac{1}{\eta_{0, \mu}({\bf x})} = \frac{1}{\eta_{0, \mu}^{(0)}({\bf x})} - \hbar^2 \frac{\eta_{0, \mu}^{(2)}({\bf x})}{(\eta_{0, \mu}^{(0)}({\bf x}))^2} + o(\hbar^2).
\]
If we assume that such relations hold, even out of equilibrium, the definition of the local temperature can be given in terms of the Lagrangian multipliers as follows.

\begin{definition}
The local temperature of a system of two or more branches of phonons is $T_{LE} := \frac{1}{k_B\eta^{LE}_{0}({\bf x})},$ where $\eta^{LE}_{0}({\bf x})$ is the common Lagrange multiplier that the occupation numbers of the branches, taken into account, would have if they were in the local thermodynamic equilibrium corresponding to their total energy density, that is, the following:
\begin{equation}
W(\eta^{LE}_{0}({\bf x})) :=\sum_{\mu}W_{\mu}(\eta_{0, \mu}({\bf x})) =\sum_{\mu} W_{\mu} (\eta_{0}^{LE}({\bf x})), \label{TLE}
\end{equation}
where the sum runs over all the phonon branches.
\end{definition}

At global equilibrium the temperature is constant $T= \bar{T}$ and the Wigner function reduced to the Bose -Einstein distribution
\begin{eqnarray}
& & g_{\mu}  = \left[\displaystyle{e^{h_{\mu}({\bf q})/k_B \bar{T}} } - 1\right]^{-1}. 
\end{eqnarray}
with the same temperature for each phonon branch.  

Let us consider a small perturbation $\delta T ({\bf x})$ of the temperature in the sense that $\delta ({\bf x})/\bar{T} \ll 1$.
We can expand $g^{MEP}_{\mu}$ in powers of $\delta ({\bf x})/\bar{T}$
\begin{eqnarray*}
g^{MEP}_{\mu} &=& \left[\displaystyle{e^{h_{\mu}({\bf q})/k_B \bar{T}} } - 1\right]^{-1} + \left[\displaystyle{e^{h_{\mu}({\bf q})/k_B \bar{T}} } - 1\right]^{-2} e^{h_{\mu}({\bf q})/k_B \bar{T}} 
\frac{h_{\mu}({\bf q})}{k_B \bar{T}} 
\frac{\delta T({\bf x})}{\bar{T}}\\
 & & + \hbar^2 \bar{T} \frac{\partial g_{2, \mu}^{MEP} (\bar{T})}{\partial T} \frac{\delta T}{\bar{T}} + o \left(\hbar^2 \frac{\delta T}{\bar{T}}\right).
\end{eqnarray*}
We observe that typically the relaxation energy relaxation time is much longer than the energy-flux relaxation times, that is $\tau^{\bf Q}  \ll \tau^W$.   
In a long time scaling, much longer than $\tau^{\bf Q}$, we get
\begin{equation}
{\bf Q}_{\mu} = - \tau^{\bf Q} \left[
\nabla_{\bf x}{\bf J}_{\mu} +\frac{\hbar^2}{(2\pi)^d}\frac{\partial^3}{\partial x_i\partial x_j \partial x_k}\int_{\mathbb{R}^d}{\bf c}\frac{h ({\bf q})}{24} g_{0, \mu}^{MEP} (\bm{\eta}^{(0)})  \frac{\partial^3}{\partial q_i\partial q_j \partial q_k} h({\bf q})d{\bf q} \right]. \label{Qasymp}
\end{equation}
The relation between the Lagrange multipliers and the basic fields, as seen, can hardly be inverted analytically but a numerical procedure is necessary. However, if we consider a situation where the system is not too far from the equilibrium an expansion of the Lagrange multipliers around the equilibrium state can be performed.  At equilibrium  $g^{MEP}$ is isotropic and therefore ${\bm \eta}_1^{equil} = {\bf 0}$ and in a neighborhood of the equilibrium ${\bm \eta}_1$ remains {\it small}. Therefore, for small deviations from the thermodynamic equilibrium the expansion
$$
g^{(0) MEP}_{\mu} = \left[\displaystyle{e^{h_{\mu}({\bf q})/k_B T}} - 1\right]^{-1} - \left[\displaystyle{e^{h_{\mu}({\bf q})/k_B T} } - 1\right]^{-2} e^{h_{\mu}({\bf q})/k_B T} 
 h_{\mu}({\bf q}) {\bm \eta}_{1, _{\mu}} \cdot  {\bf c}_{\mu} + O(\hbar^2).
$$
is valid.

By substituting in (\ref{Qasymp}) one gets up to first order in $\hbar^2$
\begin{eqnarray*}
&&{\bf Q}_{\mu} =  
-\tau^{\bf Q}
\nabla_{\bf x}{\bf J}_{\mu} \\
&&- \tau^{\bf Q} \frac{\hbar^2}{(2\pi)^d}\frac{\partial^3}{\partial x_i\partial x_j \partial x_k}\int_{\mathbb{R}^d}{\bf c}\frac{h_{\mu}({\bf q})}{24} 
\left[\displaystyle{e^{h_{\mu}({\bf q})/k_B T} } - 1\right]^{-2} e^{h_{\mu}({\bf q})/k_B T} 
 h_{\mu}({\bf q}) {\bm \eta}_{1, _{\mu}} \cdot  {\bf c} 
\frac{\partial^3}{\partial q_i\partial q_j \partial q_k} h_{\mu}({\bf q}) d{\bf q}.
\end{eqnarray*}
In particular, at the zero order we have
\begin{eqnarray*}
{\bf Q}^{(0)}_{\mu} = - \tau^{\bf Q} \nabla_{\bf x}{\bf J}^{(0)}_{\mu}
= - \frac{\tau}{(2\pi)^d} \nabla_{\bf x} \int_{\mathbb{R}^d}{\bf c}_{\mu}\otimes{\bf c}_{\mu} h_{\mu} ({\bf q})g^{(0) MEP}_{\mu} ({\bf x}, {\bf q},t) d{\bf q}.\\
= - \frac{\tau^{\bf Q}}{(2\pi)^d} \nabla_{\bf x} \int_{\mathbb{R}^d}{\bf c}_{\mu}\otimes{\bf c}_{\mu} h_{\mu} ({\bf q})  \left[\displaystyle{e^{h_{\mu}({\bf q})/k_B T}} - 1\right]^{-1} d{\bf q}\\
= - \frac{\tau^{\bf Q}}{(2\pi)^d} \int_{\mathbb{R}^d}{\bf c}_{\mu}\otimes{\bf c}_{\mu} h_{\mu} ({\bf q}) \frac{\partial}{\partial T} \left[\displaystyle{e^{h_{\mu}({\bf q})/k_B T}} - 1\right]^{-1} d{\bf q} \, \nabla_{\bf x} T_{\mu}\\
= - \frac{\tau^{\bf Q}}{(2\pi)^d k_B T^2} \int_{\mathbb{R}^d}{\bf c}_{\mu}\otimes{\bf c}_{\mu} h_{\mu}^2 ({\bf q})  \frac{e^{h_{\mu}({\bf q})/k_B T}}{\left(e^{h_{\mu}({\bf q})/k_B T} - 1\right)^{2}} d{\bf q} \, \nabla_{\bf x} T_{\mu}
\end{eqnarray*}
which can be written in the Fourier form
$$
{\bf Q}^{(0)}_{\mu} = - {\bf K}_{\mu}^{(0)} \nabla_{\bf x} T_{\mu}
$$
with the thermal conductivity tensor given by
$$
{\bf K}_{\mu}^{(0)} = \frac{\tau^{\bf Q}}{(2\pi)^d k_B T^2} \int_{\mathbb{R}^d}{\bf c}_{\mu}\otimes{\bf c}_{\mu} h_{\mu}^2 ({\bf q}) \frac{e^{h_{\mu}({\bf q})/k_B T}}{\left(e^{h_{\mu}({\bf q})/k_B T} - 1\right)^{2}} d{\bf q}. $$
It is evident that ${\bf K}_{\mu}$ is positive definite. 

Observe that $\forall {\bf n} \in {S^d}$
$$
\int_{S^d} n_{i_1} n_{i_2} \cdots n_{i_r} d \Omega = 0\quad \mbox{if } r \quad \mbox{odd},
$$
$S^d$ being the unit sphere in $\mathbb{R}^d$. Therefore, if  $h_{\mu} ({\bf q})$ is isotropic  ${\bf K}_{\mu}$ is isotropic as well
$$
{\bf K}_{\mu}^{(0)} = \frac1d k^{(0)} {\bf I}, 
$$ 
with ${\bf I}$ identity matrix of order $d$ and $k^{(0)}$ the zero order trace 
$$ 
k^{(0)} = \frac{\tau^{\bf Q}}{(2\pi)^d k_B T^2} \int_{\mathbb{R}^d} \left| {\bf c}_{\mu} \right|^2  h_{\mu}^2 ({\bf q}) \frac{e^{h_{\mu}({\bf q})/k_B T}}{\left(e^{h_{\mu}({\bf q})/k_B T} - 1\right)^{2}} d{\bf q}.$$
The second order correction in $\hbar^2$ reads
\begin{eqnarray*}
& & {\bf Q}^{(2)}_{\mu} = - \frac{\tau^{\bf Q}}{(2\pi)^d}  \nabla_{\bf x} \int_{\mathbb{R}^d}{\bf c}_{\mu}\otimes{\bf c}_{\mu} h_{\mu} ({\bf q})g_{\mu}^{(2)} ({\bm \eta}^{(0)}({\bf x}, {\bf q},t)) d{\bf q} \\
& & -  \frac{\tau^{\bf Q}}{(2\pi)^d}\frac{\partial^3}{\partial x_i\partial x_j \partial x_k}\int_{\mathbb{R}^d}{\bf c}\frac{h_{\mu}({\bf q})}{24} 
\left[\displaystyle{e^{h_{\mu}({\bf q})/k_B T} } - 1\right]^{-2} e^{h_{\mu}({\bf q})/k_B T} 
 h_{\mu}({\bf q}) {\bm \eta}_{1, _{\mu}} \cdot  {\bf c} 
\frac{\partial^3}{\partial q_i\partial q_j \partial q_k} h_{\mu}({\bf q}) d{\bf q}. 
\end{eqnarray*}
Indeed the last term in the previous relation is of order $\hbar^2 \dfrac{\delta T}{T}$ and can be considered negligible for small deviations from local equilibrium. The remaining part gives a highly nonlinear correction which cannot be put in a Fourier form. 

As an example we consider the case of the longitudinal and transversal acoustic phonons in the Debye approximation for a single branch. In such a case
the corresponding symbol of the phonon hamiltonian reads $c |{\bf q}|$ and therefore
\begin{eqnarray}
{k}_{ac}^{(0)} &=& \frac{\tau^{\bf Q}}{(2\pi)^d k_B T_{ac}^2} \int_{\mathbb{R}^d} c^4  |{\bf q}|^2  \frac{e^{c |{\bf q}|/k_B T_{ac}}}{\left(e^{c |{\bf q}|/k_B T_{ac}} - 1\right)^{2}} d{\bf q} \nonumber \\
&=& \frac{\tau^{\bf Q} c^4}{(2\pi)^d k_B T_{ac}^2} 
\mbox{mis}(S_d)
\int_0^{+ \infty}   |{\bf q}|^{d +1}  \frac{e^{c |{\bf q}|/k_B T_{ac}}}{\left(e^{c |{\bf q}|/k_B T_{ac}} - 1\right)^{2}}  d |{\bf q}| \nonumber 
\\ &=&
 \frac{k_B \tau^{\bf Q} c^{3-d}}{(2\pi)^d} 
\mbox{mis}(S_d) \left(k_B T_{ac} \right)^{d-1} 
\int_0^{+ \infty}  z^{d+1} \frac{e^{z}}{\left(e^z - 1\right)^{2}}  d \, z
\end{eqnarray}
where
\begin{eqnarray*}
\mbox{mis} (S_d)=\frac{2 \pi^{d/2}}{\Gamma (d/2)}
\end{eqnarray*}
is the measure of $S_d$,
 $\Gamma (x)$ being the Euler gamma function. 
The previous integral is convergent for any $d\in \mathbb{N}$. Observe that we get a dependence on the temperature proportional to $T_{ac}^{d -1}$.
Regarding the second order correction we observe that
\begin{eqnarray*}
g^{MEP}_{2} =-\frac{1}{8}\frac{e^\xi}{(e^\xi-1)^3}\left\{\frac{c^2(e^\xi+1)}{k_B^2T(\textbf{x},t)^4|\textbf{q}|^2}\left[\delta_{ij}|\textbf{q}|^2\left(2\frac{\partial T}{\partial x_i}\frac{\partial T}{\partial x_j}-T\frac{\partial^2 T}{\partial x_i\partial x_j}\right) \right.\right.\\
\left.\left. +q_iq_j\left(T\frac{\partial^2 T}{\partial x_i\partial x_j}-3\frac{\partial T}{\partial x_i}\frac{\partial T}{\partial x_j}\right)\right]\right.\\
\left.  -\frac{c^3(e^{2\xi}+4e^\xi+1)}{3k_B^3|\textbf{q}|(e^\xi-1)T(\textbf{x},t)^5}\left[(\delta_{ij}|\textbf{q}|^2-q_iq_j)\frac{\partial T}{\partial x_i}\frac{\partial T}{\partial x_j}-q_iq_jT\frac{\partial^2 T}{\partial x_i\partial x_j}\right]\right\}\\
 = -\frac{1}{8}\frac{c^2 e^\xi}{(e^\xi-1)^3}\left\{\frac{(e^\xi+1)}{k_B^2T(\textbf{x},t)^4}\left[2 |\nabla_{\bf x} T|^2 - T \Delta_{\bf x} T +  n_i n_j\left(T\frac{\partial^2 T}{\partial x_i\partial x_j}-3\frac{\partial T}{\partial x_i}\frac{\partial T}{\partial x_j}\right)\right]\right.\\
\left.-\frac{c (e^{2\xi}+4e^\xi+1) |\textbf{q}|}{3k_B^3 (e^\xi-1)T(\textbf{x},t)^5 }\left[ (\delta_{ij} - n_i n_j)\frac{\partial T}{\partial x_i}\frac{\partial T}{\partial x_j} - n_i n_jT\frac{\partial^2 T}{\partial x_i\partial x_j} \right]\right\} 
\end{eqnarray*}
with now $\xi=c|{\bf q}|/k_B T$. Therefore, the second order correction to the heat flux is given by 
\begin{eqnarray*}
 {\bf Q}^{(2)}_{\mu} = - \tau^{\bf Q} \nabla_{\bf x} {\bf J}^{(2)}_{\mu}
\end{eqnarray*}
with
\begin{eqnarray*}
{\bf J}^{(2)} = \frac{1}{(2\pi)^d} \int_{\mathbb{R}^d}{\bf c}\otimes{\bf c} h ({\bf q}) g^{MEP}_{2} d{\bf q} = 
\frac{c^2}{(2\pi)^d} \int_{\mathbb{R}^d} n_h n_k  h ({\bf q}) g^{MEP}_{2} d{\bf q} \, {\bf e_h}\otimes{\bf e_k} := {\bf J}^{(2)}_{hk} {\bf e_h}\otimes{\bf e_k} 
\end{eqnarray*}
$( {\bf e_1}, {\bf e_2}, \cdots,  {\bf e_d})$ being an orthonormal basis of $\mathbb{R}^d$.

By taking into account the well-known formulas
\begin{eqnarray*}
& &\int_{\Omega} n_h n_k d \Omega = \frac{\mbox{mis} (S_d)}{d}  \delta_{ij}, \\
& & \int_{\Omega} n_i n_j n_h n_k d \Omega = \frac{\mbox{mis} (S_d)}{d(d+2)}  (\delta_{ij} \delta_{hk} + \delta_{ih} \delta_{jk}+ \delta_{ik} \delta_{jk})
\end{eqnarray*}
and that
\begin{eqnarray*}
&&\int_{0}^{+\infty} \frac{h({\bf q}) e^{\xi} (e^{\xi}+1 )}{(e^{\xi}-1)^3} q^{d-1} dq = c \left( \frac{k_B T}{c} \right)^{d+1} \int_{0}^{+\infty} \frac{ e^{\xi} (e^{\xi}+1 )}{(e^{\xi}-1)^3} \xi^d d \xi := 
c \left( \frac{k_B T}{c} \right)^{d+1} I_1 (d), \\
&&\int_0^{+\infty} \frac{ h({\bf q}) e^{\xi} (e^{2\xi}+4e^{\xi}+1)}{(e^{\xi}-1)^4} q^d dq = c \left( \frac{k_B T}{c} \right)^{d+2} \int_0^{+\infty} \frac{ e^{\xi} (e^{2\xi}+4e^{\xi}+1)}{(e^{\xi}-1)^4} \xi^{d+1} d \xi \\
&&:= c \left( \frac{k_B T}{c} \right)^{d+2} I_2(d),
\end{eqnarray*}
the components of ${\bf J}^{(2)}$ read


\begin{eqnarray*}
&&{\bf J}^{(2)}_{hk}=- \frac{c^3}{8(2\pi)^d}  \frac{\mbox{mis}(S_d)}{d}  \frac{1}{k_B^2 T^4({\bf x},t)} \left( \frac{k_B T}{c}  \right)^{d+1} \\
&&\left\{  \left[ ( 2 |\nabla_x T|^2 - T \Delta_x T)   I_1(d) -\frac{1}{3}  \frac{\partial T}{\partial x_i}  \frac{\partial T}{\partial x_j} \delta_{ij}  I_2(d) \right] \delta_{hk}  
+ \left[ \left( T\frac{\partial^2 T}{\partial x_i \partial x_j } - 3 \frac{\partial T}{\partial x_i}  \frac{\partial T}{\partial x_j} \right) I_1(d) \right. \right. \\ 
&& +\left. \left. \frac{1}{3} \left( \frac{\partial T}{\partial x_i}  \frac{\partial T}{\partial x_j} +  T\frac{\partial^2 T}{\partial x_i \partial x_j } \right)  I_2(d)  \right] (\delta_{ij} \delta_{hk} + \delta_{ih} \delta_{jk}+ \delta_{ik} \delta_{jk})   \right\}.
\end{eqnarray*}


The integrals $I_1(d)$ and $I_2(d)$ are divergent in the cases $d =1$ and $d =2$. As a consequence, the quantum corrections are valid only in the bulk ($d =3$) case where $
I_1(3) = \pi^2,  
I_2(3) = 4\pi^2$. This peculiarity is physically related to the density of states and the form of the energy dispersion relations. 

\section*{Conclusions and acknowledgements}
The Wigner equation for phonons has been written in the case of  a generic dispersion relation. Moment equations have been deduced and closed by QMEP. Under a long-time scaling an expression for the heat flux with a nonlinear quantum correction has been obtained. The model is suited for the investigation in modern micro-devices where the enhanced miniaturization makes thermal effects more and more relevant.  

The authors acknowledge the support from INdAM (GNFM) and from Universit\`a degli Studi di Catania,  Piano della Ricerca 2020/2022 Linea di intervento 2 ''QICT'', V. D. Camiola acknowledges the financial support from the project AIM, Mobilit\`a dei Ricercatori Asse I del PON R \& I 2014-2020, proposta AIM1893589.

\section*{Declarations}
\subsection*{Conflicts of interest/Competing interests}
\noindent
The authors declare they have no financial interests.
\subsection*{Data availability}
\noindent
Data sharing is not applicable to this article as no new data were created or analyzed in this study.






\end{document}